\documentclass{assets}

\usepackage{balance}  
\usepackage{graphicx} 
\usepackage{times}    
\usepackage{url}      
\usepackage[T1]{fontenc}
\usepackage{textcomp}
\usepackage[scaled=.92]{helvet} 
\usepackage{booktabs} 
\usepackage{ccicons}  
\usepackage{ragged2e} 
\usepackage{etoolbox}
\usepackage{xcolor}
\usepackage{xparse}
\usepackage{listings}
\usepackage{xspace}

\begin{document}

\setcopyright{acmcopyright}
\doi{10.475/123_4}
\isbn{123-4567-24-567/08/06}
\conferenceinfo{ASSETS '17}{October 29--November 1, 2017, Baltimore, MD, USA}
\acmPrice{\$15.00}

\newtoggle{highlightchanges}
\togglefalse{highlightchanges}

\iftoggle{highlightchanges}{
   \newcommand {\changes}[1]{{\color{purple}{#1}\normalfont}}
}{
  \newcommand {\changes}[1]{{#1}}
}

\newcommand{\tool}{FluxMarker}

\title{\tool{}: Enhancing Tactile Graphics\\ with Dynamic Tactile Markers}

\numberofauthors{1}
\author{%
  \alignauthor{
    Ryo Suzuki$^1$,
    Abigale Stangl$^2$,
    Mark D. Gross$^2$,
    Tom Yeh$^1$ \\
    \affaddr{
        $^1$Department of Computer Science, $^2$ATLAS Institute
    } \\
    \affaddr{
        University of Colorado Boulder
    } \\
    \affaddr{
        Boulder, CO, USA
    } \\
    \affaddr{%
      \{ryo.suzuki, abigale.stangl, mdgross, tom.yeh\}@colorado.edu
    }
  }
}

\maketitle

\begin{abstract}
For people with visual impairments, tactile graphics are an important means to learn and explore information. 
However, raised line tactile graphics created with traditional materials such as embossing are static. While available refreshable displays can dynamically change the content, they are still too expensive for many users, and are limited in size. These factors limit wide-spread adoption and the representation of large graphics or data sets. In this paper, we present FluxMaker, an inexpensive scalable system that renders dynamic information on top of static tactile graphics with movable tactile markers.
These dynamic tactile markers can be easily reconfigured and used to annotate static raised line tactile graphics, including maps, graphs, and diagrams. We developed a hardware prototype that actuates magnetic tactile markers driven by low-cost and scalable electromagnetic coil arrays, which can be fabricated with standard printed circuit board manufacturing. We evaluate our prototype with six participants with visual impairments and found positive results across four application areas: location finding or navigating on tactile maps, data analysis, and physicalization, feature identification for tactile graphics, and drawing support. The user study confirms advantages in application domains such as education and data exploration.
\end{abstract}

%
%

%
%


\begin{figure}[t]
\centering
\includegraphics[width=3.3in]{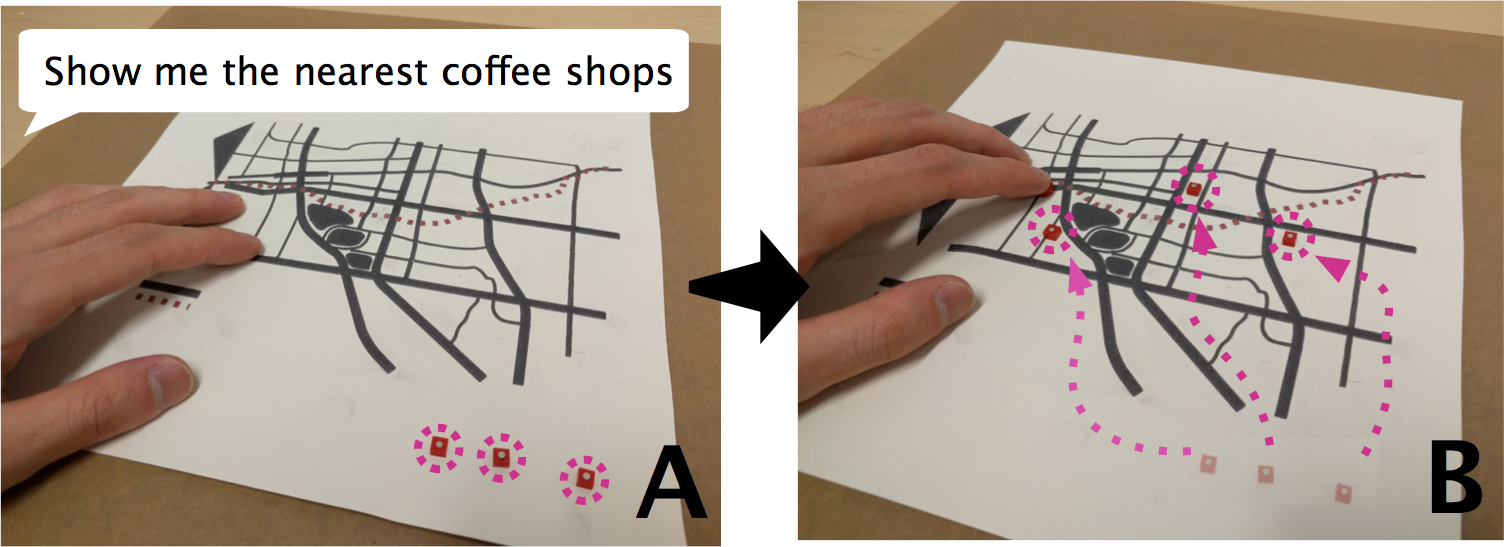}
\caption{FluxMarker User Interaction with Dynamic Markers }
~\label{fig:system-concept}
\vspace{-0.35in}
\end{figure}

\section{Introduction}
For people who are blind or visually impaired, tactile graphics are essential resources to learn and explore non-textual information such as maps, graphs, and pictorial diagrams. 
Students with visual impairments often use tactile graphics to understand an abstract concept in physics, the structure of molecules in chemistry, or a human brain model in biology classes.
Although such tactile representations are useful in presenting spatial and visual information, the form to associate the visual and textual information has been limited;
tactile graphics can only present a fixed and limited amount of information to tactile readers~\cite{brock2015interactivity}. 
Braille labels and keys are often used to annotate the content~\cite{BANA}, but this annotation is not accessible for many blind users, as approximately 90 percent of blind people in the United States cannot read braille~\cite{national2009braille, blindstatics2015}.

To overcome these limitations, recent work has focused on enhancing traditional tactile graphics with interactive tactile graphics, which leverage multimodal interaction to provide more accessible and adaptable information associated with the visual information.
These interactive graphics can detect touch input and annotate content with an audio description~\cite{baker2014tactile, miele2006talking} or haptic feedback~\cite{yu2001haptic}, which allows readers to explore the tactile content more efficiently than classical tactile maps or diagrams~\cite{brock2015interactivity, landau2003merging}.
However, the tactile representations of these systems are still static and they cannot render dynamic content in tactile form~\cite{jacobson1998navigating}.
Through our formative study, we found that the current way to present dynamic tactile representations is largely limited with small-size and expensive refreshable braille displays, which can significantly limit both potential applications and wide-spread adoption.

This paper proposes {\it dynamic tactile markers}, a new approach to enhancing tactile graphics with reconfigurable multiple tactile elements.
Dynamic tactile markers are movable, self-adjusting physical elements that can render the dynamic information on top of a traditional tactile graphic such as swell paper or thermoformed plastic.
In contrast to prior approaches, which utilize audio or haptic feedback, our approach aims to enhance {\it tactile} feedback which can provide the real-time affordances and the physical guides for blind or low visions users to explore the spatial information.

We explore four application scenarios where the dynamic tactile markers assist blind users accomplish the following tasks: (1) find locations on a map, (2) read and analyze dynamic data, (3) locate and identify specific features on tactile graphics, and (4) draw through dynamic assistance. In these scenarios, the dynamic tactile markers can be used as a tangible data point for data visualization, a location or navigating path on a map (Figure~\ref{fig:system-concept}), or spatial reference points for guided drawings.

This approach is motivated by a formative study in which we asked four blind participants the needs and challenges of the current tactile representations. 
From the study we learned that high cost and small working space size are the major limitations for accessing the technology, which suggests an opportunity for a new type of interaction with tactile graphics.
Based on these findings and the consideration of different actuation techniques, we explore an electromagnetic actuation as a low-cost and scalable design for enhancing existing tactile graphics with dynamic tactile markers.

To demonstrate this concept, we present FluxMarker, a software and hardware prototype that actuates magnetic elements on top of a static tactile graphic. \changes{FluxMarker can move multiple small magnets to a grid of possible locations by using an array of electromagnetic coils.} The coils are fabricated with standard printed circuit board (PCB) manufacturing techniques, which can enable the low-cost fabrication (40 USD for a 16x16 grid and 15cm x 15cm dimension, and 500 USD for a 160x160 grid and 150cm x 150cm dimension). With modular design, the size of the display easily scales up without significant increase of cost and fabrication complexity, while allowing independent multiple magnets control.

We evaluated our prototype with six people with visual impairments to investigate the plausibility of the application scenarios identified during our formative work. 
We found that all participants were able to use the FluxMarker to identify specific features on the tactile graphics faster than when they did not have a reference point, albeit they wanted to have the markers move along paths to guide them between landmarks. They were also interested in using the markers to create raised lines around specific tactile elements so that they could feel the boundaries and the contained tactile information. Our participants also noted the possibility for the system to annotate graphics in real-time, which would help understand their data sets, interpret tactile graphics at the same time as teachers present the same information visually during lectures, and with building in situ ways to navigate. Finally, our participants confirmed that the FluxMarker would help people learn to draw, in particular, young students.

In summary, our contributions are as follows: 
\begin{itemize}
\item An approach to enhancing the tactile graphics with dynamic tactile markers
\item A design of low-cost, scalable actuated tangible markers, informed by a formative study with four blind people.
\item A hardware prototype of the PCB manufactured electromagnetic coils and its technical evaluation.
\item A user evaluation study with four blind participants and two low vision participants, which illustrates the potential benefits of dynamic tactile markers in four application scenarios. 
\end{itemize}

\section{Related Work}

\subsection{Interactive Tactile Graphics}
Although the benefits of tactile graphics are well documented, there are several limitations. 
The first limitation of a tactile graphic is its finite capacity to hold information~\cite{brock2015interactivity}. It is difficult to add information, such as captions or annotations, without making a tactile graphic overly complicated~\cite{tatham1991design}. Take for example a tactile map with roads, intersections, and several landmarks. It would not be feasible to add a tactile label to every map feature. Researchers have been exploring the use of other modalities to augment a tactile graphic with additional information. Two of the most promising modalities are sound~\cite{baker2014tactile, miele2006talking} and haptics~\cite{yu2001haptic}. 
Sound has been applied to annotate the content of a tactile graphic to give a text-to-speech description based on QR code~\cite{baker2014tactile}, object recognition~\cite{fusco2015tactile}, or touch input~\cite{miele2006talking}. 
Haptic-tactile maps~\cite{rice2005design, zeng2010audio} can generate force feedback based on the user interaction. 
Compared with traditional tactile graphics these interactive tactile graphics can improve the efficiency in exploring content and facilitate learning ~\cite{brock2015interactivity, landau2003merging}. 
However, sound and haptics have their own limitations:
they limit users' ability to obtain quick overviews of spatial information with two handed interaction~\cite{mcgookin2010clutching} and using the hands as a marking or reference points to compare different parts of the graphic spatially~\cite{rice2005design}.
In contrast, our proposed dynamic markers are designed to improve {\it tactile} feedback by providing physical guides and affordances directly on a tactile graphic for blind users to explore and comprehend spatial information.

\subsection{Dynamic Tactile Graphics}
Another limitation of a tactile graphic is its static content and the high cost of production~\cite{jacobson1998navigating}.
Although recent work has demonstrated tools to automate the design of tactile graphics~\cite{brown2012viztouch, jayant2007automated, vstampach2016automated}, once it is created a static tactile graphic cannot be easily modified.
A dynamic tactile graphic can enable updating of its content in response to users' inputs. HyperBraille~\cite{prescher2010tactile} is a commercially available refreshable braille display that has one of the largest touch-sensitive pin-matrix display (7200 pins arranged in 60 rows). 
Researchers have demonstrated interactive systems that leverage such commercially available refreshable displays to produce a dynamic tactile map with geographic annotation~\cite{schmitz2012interactively, zeng2012atmap}.
However, the cost of a dynamic tactile display like HyperBraille is prohibitive, ranging from 2,000 USD for an 18-character display to 50,000 USD for a half page of braille. 

Recently, a wide variety of novel actuator technologies has been proposed, including electromagnetic actuators~\cite{yeh2007mechanism}, piezo-electric actuators~\cite{cho2006development, volkel2008tactile}, electroactive polymers~\cite{chakraborti2012compact}, hydraulic and pneumatic actuation~\cite{lee2005micromachined, russomanno2015design}, and shape memory alloy~\cite{taylor1998sixty}. 
However, piezo-electric actuators are still the only technology found in commercially available devices~\cite{russomanno2015design}, and the cost of a single piezo-powered braille cell is approximately 100 USD, bringing the cost of even a single line refreshable braille display to over 1,000 USD~\cite{runyan2010eap}. 
To enable dynamic updating of tactile content, we explore an alternative approach where instead of developing an alternative refreshable braille display we augment a tactile graphic.
Our hybrid approach allows a blind user to interact with the tactile content dynamically, while allowing size and resolution to scale without a significant increase in cost and fabrication complexity. 

\subsection{Tangible Interaction}
One emerging form of dynamic tactile graphics are those enabled by a tabletop tangible user interface.
Tabletop tangible user interfaces were first created to allow users to interact with digital information by moving or actuating physical objects~\cite{ishii1997tangible, pangaro2002actuated, patten2007mechanical}, and these systems have been applied to many domains, including urban planning~\cite{underkoffler1999urp}, remote collaboration~\cite{follmer2013inform}, education~\cite{horn2009comparing}, and data visualization~\cite{le2016zooids}.
Recently, researchers have investigated ways to use tangible interfaces for assistive applications~\cite{mcgookin2010clutching, schneider2000constructive}.
For example, Tangible Graph Builder~\cite{mcgookin2010clutching} is specifically designed for visually impaired users to allow them to access graph and chart-based data through tangible interface.
Tangible Reels~\cite{ducasse2016tangible} helps visually impaired users to construct a tangible map by their own with sucker pads and retractable reels.
These devices allow visually impaired people to dynamically create tactile maps and retrieve specific information related to points and links.
Inspired from these work, we explore how {\it actuated} tangible objects can enhance the exploration and interaction with tactile graphics for visually impaired users.

\section{Formative Study}

We conducted a formative study with four blind individuals (male: 2, female: 2) to understand their current uses of tactile graphics, challenges they encounter, and opportunities where dynamic tactile markers may be helpful for them.
\changes{The age of the participants ranged from 22 to 28 (Mean=25.75, SD=2.6).}
All participants were students (one undergraduate and three graduate students) in various fields (biology and neuroscience, astrophysics, and computer science) at a local university.
We chose students as main target users because tactile graphics are heavily used in education, particularly in STEM fields.
Throughout a 30-minute semi-structured interview, we focused on three aspects: (1) current use of tactile graphics, (2) challenges and limitations in the current use of tactile graphics, and (3) opportunities for an alternative approach to enhancing tactile representations.
Next we present our findings.

\subsection{Current Uses}
We first asked the participants when and why they use tactile graphics. All participants have used tactile graphics for their coursework or research. For example, P2 said that she uses tactile graphics to access visual material in the textbooks of her biology and neuroscience classes. She uses a tactile representation of the brain model to spatially understand the functionality of each anatomical region. A particularly important use scenario is data exploration and analysis. P4 was involved in a space grant project that sends a balloon with instruments to high attitude to collect data. P4 mentioned that a tactile graphic would be a good medium to represent the data in an accessible way for analysis.

\subsection{Challenges}
Participants reported several challenges in using a \emph{static} tactile graphic for data analysis and learning resources. First, they found it difficult to understand changes in data. P4 pointed out that this difficulty is due to the lack of dynamicness in tactile representation. In addition, when there is too much information, a graphic can be too complex to interpret\changes{~\cite{brock2015interactivity,tatham1991design}}. P2 commented that {\it ``P2: When you try to add all the information on a single tactile graphic, this can be too complex.''} 

Participants also reported several limitations in the current form of a \emph{dynamic} tactile representation. First, the current devices for displaying dynamic tactile graphics are very costly. All participants commented on this cost issue as a hindrance for wider adoption. P2 said that {\it ``P2: I don't have any of these [refreshable braille displays]. I want, but the cost of thousands of dollar is just too expensive for me.''} Second, the size is too small\changes{~\cite{swaminathan2016linespace}}. P1 mentioned that the small size of these displays makes it difficult to use for data analysis applications: {\it ``P1: braille display can show the 40 characters or maybe 80. That's about it. [...] I know there is an effort to make 4 lines or 5 lines of the braille display, but I'm not sure how successful they are. These can be very expensive.''} P3 mentioned that these [refreshable braille displays] are only designed for reading text, not showing data. {\it ``P3: It's too small and can't express, for example, weather map, or complicated graph of 5000 data point.''} 

In summary, a static tactile graphic lacks the ability to represent changes and can be overly complex, whereas a dynamic tactile graphic is costly and too small. Neither is ideal for supporting data analysis and interactive information retrieval.

\subsection{Opportunities}
The key opportunity we identified from the formative study is to consider a hybrid method that combines both static and dynamic tactile graphics. P2, who did not own a refreshable display, mentioned that {\it ``It would be cool if it can dynamically label the part or change the texture, so that it can keep the tactile graphic simple but as accurate as possible.''} In other words, only a part or a few parts of a graphic need to be dynamic, while the rest of the content remains static. 

Another opportunity worth noting is a common desire to understand changes in data analysis. P1 mentioned that {\it ``I'm actually more talking about the dynamics over time, say, ... how much snow falls over time, earthquake data, or global warming data, anything that are changing over time. I don't know what any of these look like in the real world.''}

\subsection{Design Requirements}

Our formative study inspired us to develop \tool{}, a technique for controlling a set of dynamic tactile makers to move around a static tactile graphic to support data exploration and analysis. Informed by our findings, we identified the following design requirements for \tool{}:
\begin{enumerate}
\item \textbf{Support:} It needs to support a range of traditional static tactile graphics.
\item \textbf{Dynamic Update:} It needs to dynamically update its location in response to user inputs.
\item \textbf{Multiple Markers:} It needs to be capable of controlling multiple markers independently.
\item \textbf{Perceptibility:} Its location as well as changes in location need to be perceivable by users via hands.
\item \textbf{Scalability:} Its cost needs scale well as the display area increases, preferably linearly.
\end{enumerate}

\section{Dynamic Tactile Markers}
To address the limitations of the current tactile graphics, we propose {\it dynamic tactile markers}, a new approach that uses movable self-adjusting physical elements to dynamically render points of information on top of a traditional tactile graphic. The markers are magnets that are manipulated above a bed of electromagnetic coils, whose movements are controlled by software. The magnetic, dynamic tactile markers create real-time and adjustable tactile reference points, which can easily reconfigure the tactile content and enrich the spatial information.
While dynamic tactile markers can be applied more generally in any tangible user interface, we specifically explore the design space of augmented tactile graphics for people with visual impairments. This section describes the interaction design and use scenarios that led us to investigate the design of a system to render dynamic markers.

\subsection{Interaction and Application Scenarios}
The main goal of dynamic tactile markers is to provide real-time tactile affordances on an otherwise static tactile graphic in order to direct a user's attention to specific features of the graphic.
This type of interaction is especially important when users explore spatial content.
\changes{In contrast to the refreshable braille display, the hybrid approach using the combination of a static tactile diagram and dynamic markers makes the display content persistent without losing the user's spatial memory~\cite{swaminathan2016linespace}.
This enables users to easily recognize the position of the marker by referring the static outline as a constant reference, while allowing update context-dependent contents based on the user's needs, such as a location in a map or data points of a graph.}
Here we describe four application scenarios where dynamic tactile markers can be useful for people with visual impairments.

\subsubsection{Location Finding and Feature Identification}
Tactile maps provide blind people a means to explore geographical information.
For example, a tactile map of a campus will display a layout of buildings and braille labels associated with each building. 
However, finding a particular location is often a tedious task;
unlike a sighted person's ability to scan a map and quickly identify a specific location, blind users usually explore the map sequentially and must orient themselves to the whole graphic before finding a specific location. 
Moreover, although the information is often labeled with braille, reading braille takes time and is inaccessible for those who cannot read braille. Audio feedback can help to orient users to the name or feature of the current location, however, this technique makes it difficult to orient oneself to specific locations on the page. 

\begin{figure}[h!]
\centering
\includegraphics[width=3in]{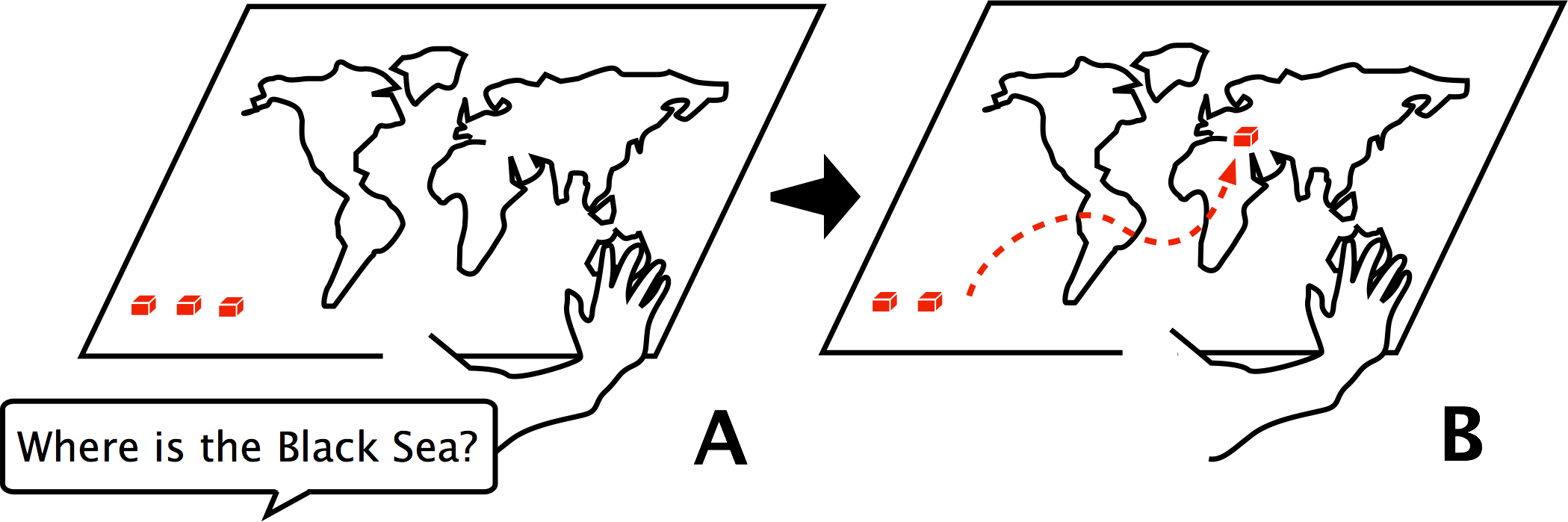}
\caption{FluxMarker's Location Finding and Feature Identification Application}
~\label{fig:application-map}
\vspace{-0.2in}
\end{figure}

Dynamic tactile markers can help to identify a spatial location quickly. For example, responding to ``Where is the nearest coffee shops?'' in a local area map or ``Where is the Black Sea?'' in a geographical world map, the dynamic tactile markers can move around on the tactile map, and the blind user can use their hands to quickly skim the map to identify the location of the marker (Figure~\ref{fig:application-map}). They can quickly find the marker position relative to their current location or an outline of surrounding areas on an existing tactile map. In this way, they do not lose contact with spatial reference points or the spatial memory they have developed. 
In addition, responding to the query, ``How can I get to this place?'', other markers can instantly draw the tactile path by aligning dots on the map. 
Once the user is satisfied by finding the location or route, the dynamic tactile markers can be reset, and cleared from the tactile graphic. 

\changes{
Similar to location finding, the dynamic tactile marker can be used to locate a specific feature on a tactile map based on a user's question. 
For example, a student with visual impairments is given a brain model to use in her biology class. She can ask ``which region of the brain has a memory function?'', the dynamic tactile marker can point out the domain of a hippocampus by positioning the marker within that region of the organ. This is a similar, but different interaction from existing interactive tactile graphics, which explains the feature of each region triggered by the user's pointing, while the dynamic tactile marker can point out the location triggered by the user's question. In another scenario, a student is in a lecture, and the professor is presenting a graphical representation of a cell via PowerPoint, and uses a laser pointer to identify the cell nucleus for the sighted students. The student with visual impairment, who has a tactile version of the graphic, can ask the dynamic tactile marker to move to the corresponding location.
}

\subsubsection{Data Analysis and Physicalization}
Data analysis is one of the most challenging tasks for people with visual impairments. As the visualized data is not accessible for blind users, they often find it difficult to interact with the data.
Dynamic tactile markers can help blind users make sense of data through data physicalization~\cite{jansen2015opportunities}.

One advantage of using dynamic tactile markers is the ability to update the data for a different context.
For example, a blind user who wants to analyze the temperature of a city over time might want to know the pattern throughout the year, maximum temperature, and minimum temperature of the city. 
Twelve dynamic tactile markers can position themselves to display a plot graph to represent the temperature data of each month.
By touching the data point and referring to the scale, which can be given by a static embossed paper, the user can find the maximum and minimum temperature of the city.
While understanding the pattern of the data can be challenging with audio representation alone, with dynamic tactile markers, she can also comprehend the pattern of the graph by recognizing spatial positions. If she wants to analyze the temperature data a different city, she can just ask ``render the data point'' with the city name.
Then, the dynamic tactile markers can be repositioned to render the requested data point.

\subsubsection{Guided Drawing Assistant}
\begin{figure}[h!]
\centering
\includegraphics[width=3in]{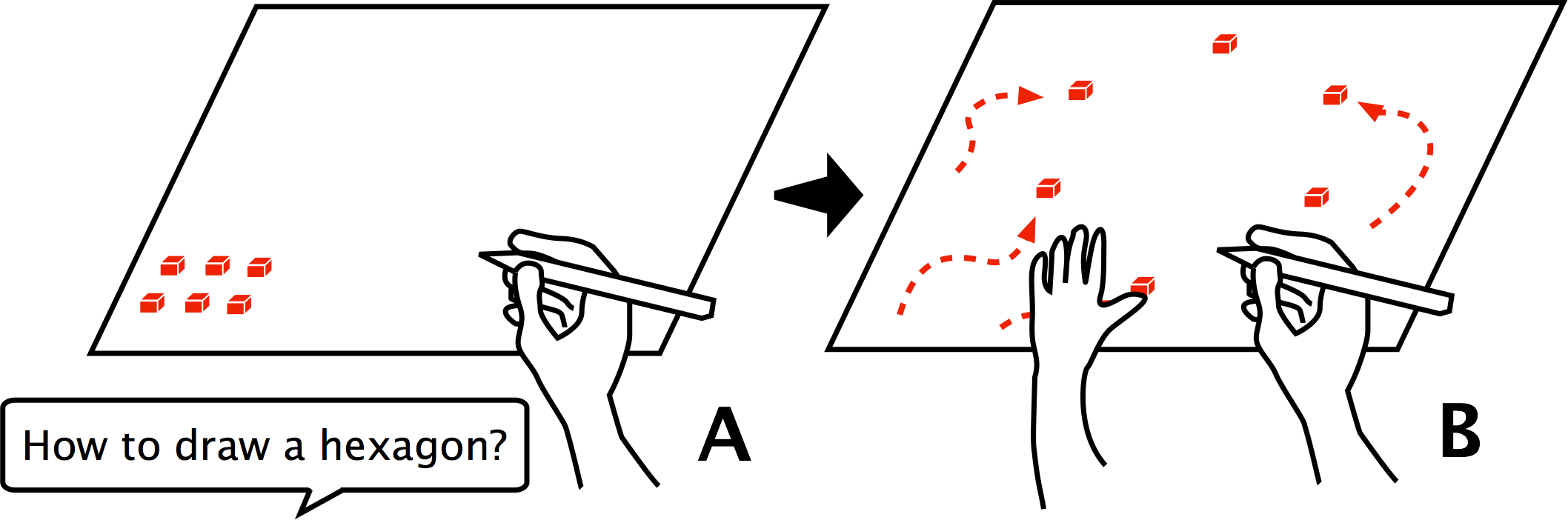}
\caption{FluxMarker's "Guided Drawing Assistant" Application.}
~\label{fig:concept-draw}
\vspace{-0.2in}
\end{figure}

In addition to supporting an interpretation of a content or analyzing the data, the dynamic tactile marker can also support students to create their own tactile graphic representations. Many students with visual impairment have limited exposure to drawing or making their own representations of information due in part to the lack of educational practices and materials~\cite{hayhoe2014reducing}. The dynamic tactile marker can help blind users to make their own tactile representations by guiding them with reference points of the drawing. For example, when a blind user is trying to draw a hexagon, six dynamic tactile markers would appear, marking the reference points of each corner of the shape. The user can touch the markers to position themselves with the nondominate hand, and guide them to draw the line to the next point (Figure~\ref{fig:concept-draw}). Or, the tactile markers can form a nearly solid edge that the user could mark alongside. This guided drawing can be particularly useful when creating their own tactile graphics when used in conjunction with inexpensive physical tactile drawing boards 
such as the Sensational Art Board~\footnote{http://www.sensationalbooks.com/products.htmlblackboard}, the inTact Sketchboard~\footnote{http://www.easytactilegraphics.com/} and 3D printing Doodle Pens~\footnote{http://the3doodler.com/}.

\section{Design Considerations}
Many different actuation techniques can enable dynamic tactile markers, but an appropriate design should meet the design requirements that we identified through our formative study. 
In order to ensure that our design meets these requirements, we evaluated a variety of actuation approaches that have been proposed in different areas such as tangible user interface, robotics, and accessibility.
These actuation methods include mechanical actuators (e.g., DC motors, servo motors, stepper motors), piezoelectric actuators (e.g., piezo-elastomer, piezo-electric linear motor, ultrasonic motor), electrostatic actuation, magnetic actuation, electromagnetic actuation, pneumatic and hydraulic actuation, and material-based actuation (e.g. shape memory alloy).
Given the technical considerations, we decided to explore electromagnetic coils to actuate a passive magnet as a marker.
Three primary considerations rose to the surface while conducting this evaluation: cost, scalability, fabrication complexity, and compliance. 
This section describes the design rationale behind our decision.

\subsection{Cost}
One of the most important considerations is the cost of fabrication. 
Although mechanical actuation such as motors and linear actuators is the straightforward design choice, these parts are expensive.
For example, coordinated self-positioning robots like Zooids system~\cite{le2016zooids} can be used as dynamic tactile markers, but parts and assembly costs 50 USD for each robot so it is costly to increase the number of markers. 
In contrast, components that can be fabricated with existing PCB manufacturing technique are inexpensive~\cite{strasnick2017shiftio},
for example, electromagnetic~\cite{pelrine2012diamagnetically, strasnick2017shiftio} or electrostatic actuation~\cite{karagozler2009stress}.
Piezo-electric actuation such as ultrasonic motors and piezo-electric linear actuators can be also integrated with PCB board~\footnote{http://pcbmotor.com/}, but the fabrication process requires piezoceramic materials and specialized manufacturing process, which increases the cost of fabrication. 
Another low-cost actuation method is pneumatic or hydraulic actuation as the parts are relatively inexpensive.

\subsection{Size and Scalability}
As we found through the formative interviews, display size is another important consideration. 
Existing approaches that use one actuator for each pixel of a dynamic tactile display, including refreshable braille~\cite{hyperbraille} or raised-pin displays~\cite{follmer2013inform, leithinger2010relief, poupyrev2004lumen}, do not scale well.
For example, a 10x10 pin size raised-pin display that uses either mechanical or piezo-electric linear actuation requires only 100 actuators.
However, a 100x100 display size requires 10,000 individually actuated pins.
Even with relatively low-cost actuators, cost increases exponentially with display size (e.g., using a 5 USD servo motor, a 100x100 pixel display will cost at least 50,000 USD).

In contrast, PCB manufactured electromagnetic actuation scales relatively well because many coils can be aligned on a PCB.
For example, in our design an 8x8 array of coils can be aligned on a 10cm x 10cm PCB, costing only 0.50 USD.
While the cost of printed circuit board increasees with the size of the board, the cost increase is trivial, and the cost of transistors to drive a high current for electromagnetic actuation is also inexpensive compared to mechanical or piezo-electric components.

\subsection{Fabrication Complexity}
In addition to cost and scalability, we value the simplicity of fabrication and control mechanism which allows the larger accessibility community to quickly adapt, replicate, and test.
As mentioned above, pneumatic and hydraulic actuation methods are also promising approaches. Researchers have proposed using a fluidic logic circuit to switch the pressure of pneumatic actuators and control the state of each pixel in a refreshable braille display~\cite{russomanno2015design}. The complexity of design and fabrication of hydraulic actuation can be alleviated with advanced 3D printing technology~\cite{maccurdy2016printable}, but it is still difficult to design complex fluidic circuits that can control the multiple pixels individually. In short, the fabrication and control mechanisms of such pnuematic actuated devices are a challenge.

In contrast, using electromagnetic coils leverages commercially available PCB manufacturing for fabrication, and a standard circuit design for the control mechanism.
Thus, we chose to develop an electromagnetic actuation technique that meets all the considerations we identified previously, while allowing the simple control and fabrication process. 


\section{System Design \& Implementation}

\begin{figure}[h!]
\centering
\includegraphics[width=3in]{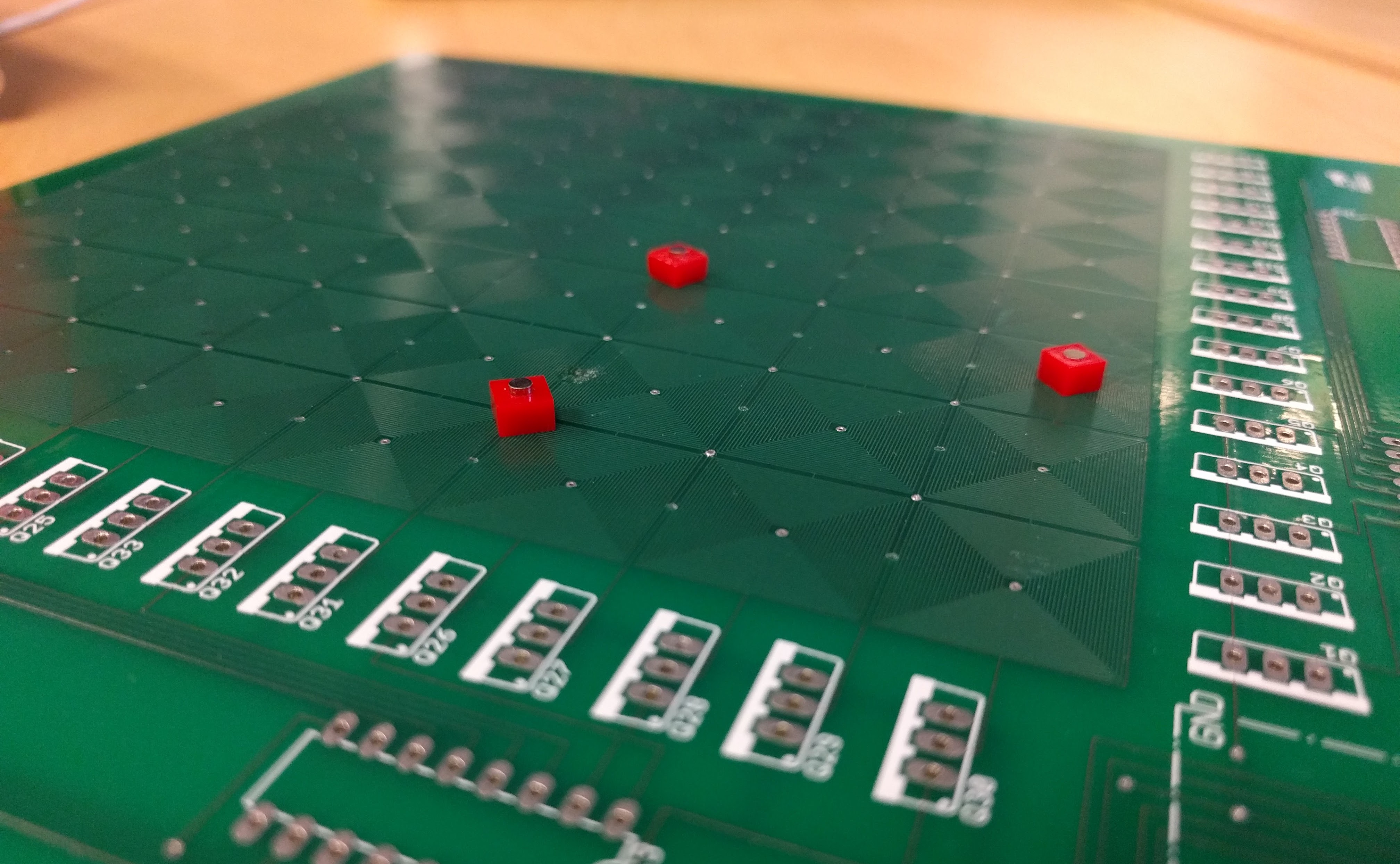}
\caption{Overview of PCB Coil Board}
~\label{fig:system-pcb}
\vspace{-0.2in}
\end{figure}

To instantiate the concept of the dynamic tactile marker, we present \tool{}, a software and hardware system that actuates magnetic markers with low-cost, scalable electromagnetic coil arrays (Figure~\ref{fig:system-pcb}). 
The hardware system is comprised of markers, coils, circuits, a controller, and a corresponding GUI. 
In the following section, we describe the specifications of the elements we used to construct the system.

\subsection{Hardware}

\subsubsection{System Design}

\textbf{Coil Design:} 
\tool{} consists of magnetic passive markers and arrays of electromagnetic coils.
The electromagnetic coil arrays can be fabricated with standard PCB manufacturing technique.
We use a two-layer printed circuit board and each layer contains a set of micro-coils with horizontal and vertical offsets.
Each coil has an identical rectangular shape and is arranged in the shape of a tile (see Figure~\ref{fig:system-pcb}).

\begin{figure}[h!]
\centering
\includegraphics[width=3in]{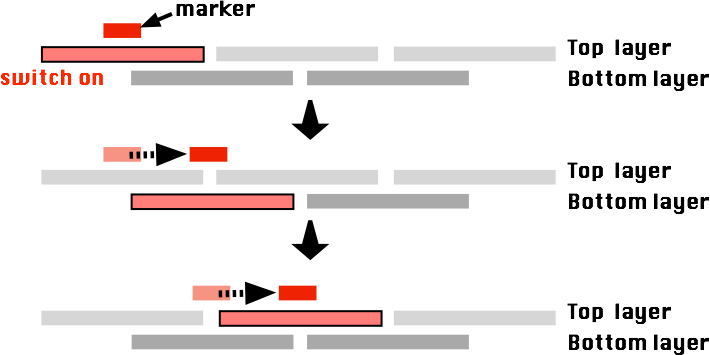}
\caption{Diagram of the Dynamic Marker's Movement Across the Coil Board}
~\label{fig:system-move}
\vspace{-0.2in}
\end{figure}

Running current through the circuit coils generates a local magnetic field within the area of the coil such that each coil can only attract a single magnet located within its area.
If the PCB had only one layer, there would be no way to move the magnet from the center of one coil to the next because the magnet is located beyond the range of the second coil.
Thus, the pattern of coils on the top and bottom layers are offset so that their effective areas overlap.
Figure~\ref{fig:system-move} illustrates the movement of the magnet. 
The microcontroller switches a sequence of coils on and off to move the magnet across the coils.
As the top layer and bottom layer are offset both horizontally and vertically, the magnet travels in a zig-zag path from one coil (on the top layer) to the next (on the bottom layer) rather than in a straight line.
\begin{figure}[h!]
\centering
\includegraphics[width=3in]{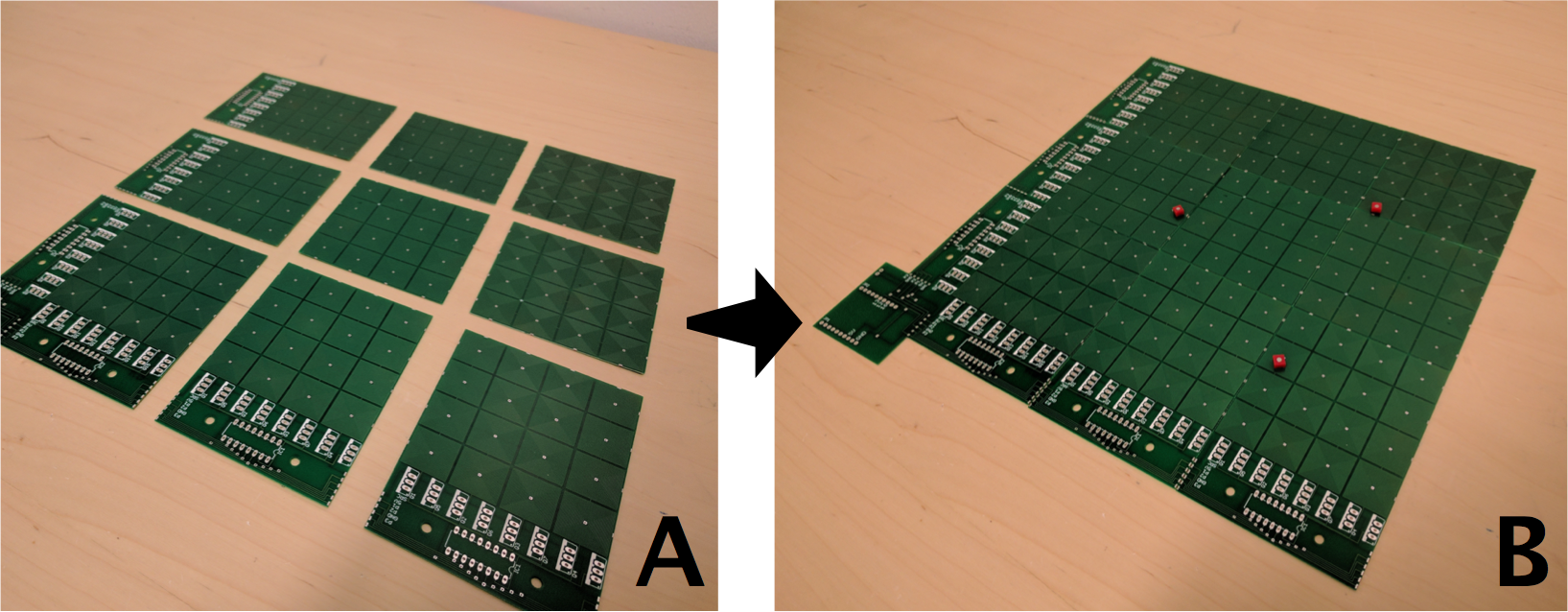}
\caption{Modular and Scalable Design}
~\label{fig:system-module}
\end{figure}

The coil arrays are fabricated with standard PCB manufacturing so the size of each array is limited by the capability of the PCB factory.
To address this, we designed our electromagnetic coil arrays as a scalable module. 
Each 16 x 16 magnetic coil array board is a module of a certain size (e.g., 15cm x 15cm).
Modular boards can be soldered together side to side as tiles, allowing the overall size of the coil array to be as large as desired (Figure~\ref{fig:system-module}).

\begin{figure}[h!]
\centering
\includegraphics[width=2in]{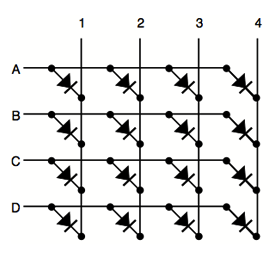}
\caption{Multiplex Coil Matrix}
~\label{fig:system-matrix}
\end{figure}

\textbf{Circuit Design:}  
Switching current to each coil turns on and off its magnetic field.
The standard approach to switching the current is to use a single MOSFET transistor for each coil, but this increases the complexity of the circuit design as it requires several of I/O lines to drive each MOSFET transistor.
Instead, we use a multiplexing technique with a diode array to moreefficiently control and drive many coils in an array.
Consider a 4x4 array of coils where each coil is connected to a diode (Figure~\ref{fig:system-matrix}). 
Similar to a LED matrix display, only one row of coils can be on at any time. By switching through each row quickly (e.g., 10-100ms), a coil at any position can be activated. For example, setting only row A as HIGH and the other rows (B, C, and D) as LOW, while setting column 1 and 3 as LOW and the other columns (2 and 4) as HIGH will turn on only coils (A, 1) and (A, 3). 
Next, if we set row B as HIGH and the other rows (A, C, and D) as LOW, and set column 1 and 4 as LOW and the other columns (2 and 3) as HIGH, we can turn on (B, 1) and (B, 4).
In this way we can control 16 coils using only 8 (4 + 4) I/O pins on the microcontroller.
This design decreases the complexity of the circuit and reduces the required number of microcontroller I/O pins as well as MOSFETs, which cost more than diodes.

While LEDs can be switched with relatively low current (e.g., 20mA) directly supplied by the microcontroller, the electromagnetic coil requires higher current (e.g., 0.5-1A).
Thus, we use half-bridge MOSFET transistor switches to amplify and control the current to each coil.
The half-bridges are made from a push-pull pair of P-channel and N-channel power MOSFET transistors.
One terminal from each coil is tied to a P-channel MOSFET transistor, and another terminal is tied to an N-channel MOSFET transistor. 
The gate of both MOSFET transistors are controlled by an I/O line from the microcontroller, and the source voltage comes from an external 9V power supply.

\textbf{Controller Design:} In this scheme, each half-bridge transistor uses two I/O pins of the microcontroller, so the number of I/O pins on the microcontroller limits the number of available transistor switches (e.g., the Arduino microcontroller has only 14 digital I/O pins).
To further reduce the required number of I/O pins, we use daisy-chained shift registers.
Each shift register switches multiple MOSFET transistors with serial-in/parallel-out data transmission. By using a chain of shift registers any number of transistors are controlled using only a few microcontroller pins.

By generating a local magnetic field, each coil attracts a magnetic marker located in its range. 
To move the marker from one point to another, the program analyzes the shortest path, and then switches coils on and off sequentially along this path to move the magnet.
As a blind or low vision user interacts with a marker, the system keeps the coil charged so that the marker cannot accidentally be pushed aside. 

\begin{figure}[h!]
\centering
\includegraphics[width=3in]{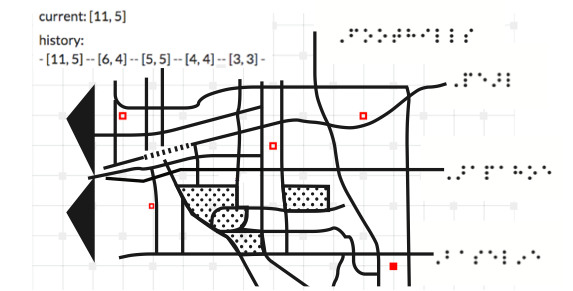}
\caption{GUI Software}
~\label{fig:system-gui}
\vspace{-0.2in}
\end{figure} 

\subsubsection{Implementation }

\textbf{Markers:} We use N50 neodymium disc magnets (2mm diameter and 3mm thickness) to act as tactile markers that are dynamically actuated by electromagnetic forces. We added a laser-cut square cap to stabilize the orientation so that the magnets do not flip over. In our prototype, each magnetic marker costs approximately 0.20 USD.

\textbf{Coils:} Our current prototype comprises 16x16 grid coils and is 15cm x 15cm in size. Each coil has 22 turns and the size is 15.85mm width and height. The width and spacing of each line in the coils is 0.1524mm (0.006 inches), the minimum trace width and separation of our PCB manufacturer, to maximize the number of turns in each coil. 

\textbf{Circuits and Microcontrollers:} We use an ESP8266 microcontroller, which switches the current to the coils using 8-bit HC595N shift registers. Each shift register can drive 8 coils which are switched using a half-bridge MOSFET transistor.
An N4001 diode attached to each coil prevents reverse current flow in the diode array. We use FQP27P06 and IRF740 for P-channel and N-channel MOSFETs, respectively. A 9V AC-DC line voltage adapter powers the MOSFETs.

\textbf{Fabrication Cost:}
Our prototype costs approximately 40 USD, including the cost of 32 MOSFETs, 128 diodes, 4 shift registers, a printed circuit board, and a microcontroller.
We estimate the cost of 160x160 grid coils will be 500 USD for the total parts cost (MOSFETs: 128 USD, diodes: 147 USD, shift registers: 20 USD, PCB: 200 USD, and microcontroller: 15 USD).
In the production of our prototype hardware system, we manually assemble these parts, but this process can automate with PCB assembly machines.

\subsection{Software}
We developed software to support the task of specifying the locations of dynamic tactile markers and controlling their movements. This software consists of a web-based graphical user interface (GUI) and a web server that communicates with the display hardware. Using our GUI, the task of creating a hybrid tactile graphic is as follows. 
First, a sighted tactile graphic designer will specify the static elements of the graphic by drawing lines and polygons or by importing an existing graphical file. 
Second, the designer will specify a spatial configuration of markers (e.g., the locations of coffee shops on a local tactile map, or the position of hippocampus in the human brain model). 
Third, the designer will specify the input commands associated with this particular configuration (e.g., voice command of ``show me the nearest coffee shops'' or ``which region of the brain is the hippocampus?''). 
Finally, if the designer wants to specify a sequence of such spatial configurations, the system supports the creation of a step-by-step guide or a drawing aid (Figure~\ref{fig:system-gui}).

The main functions of the web server component of our software are to compute the display logic given a particular marker configuration and to communicate this logic to the display hardware. The communication is through a wireless HTTP (HyperText Transfer Protocol) connection. On the hardware side, an ESP8266 microcontroller enables wireless communication with a built-in Wifi chip module. Each coil in the display matrix has a unique ID. The web server can send messages to individual coils to turn them on or off. The display logic specifies the sequence and timing of these messages in order to move markers to desired locations and the software tracks the history of each marker position. Once the task is finished, the system moves the markers to the corner of the display away from the tactile graphic. The microcontroller program is written in C++ and the control GUI is written in JavaScript.

\begin{table*}[t]
\centering
\begin{center}
\begin{tabular}{ |c|c|c|c|c|c|c| } 
 \hline
 Participant & P1 & P2 & P3 & P4 & P5 & P6 \\
 \hline
 \changes{Gender} & Male & Male & Female & Female & Male & Female \\ 
 \changes{Age} & 22 & 28 & 26 & 36 & 23 & 26 \\ 
 \changes{Visual Impairment Status} & Blind & Blind & Blind & Blind & Low Vision & Low Vision \\ 
 Frequency of Tactile Graphic & Medium & Medium & High & Medium & None & None \\ 
 Familiarity with Science Graphics & None & Medium & Medium & Medium & None & None \\
 Familiarity with Tactile Maps & High & High & Low & High & Low & Low \\
 Braille Fluency & High & High & High & Medium & Low & Low \\
 \hline
\end{tabular}
\end{center}
\caption{Characteristics and Experience of Participants in Our User Study}
~\label{fig:participant-chart}
\end{table*}

\begin{figure*}[t]
\includegraphics[width=7in]{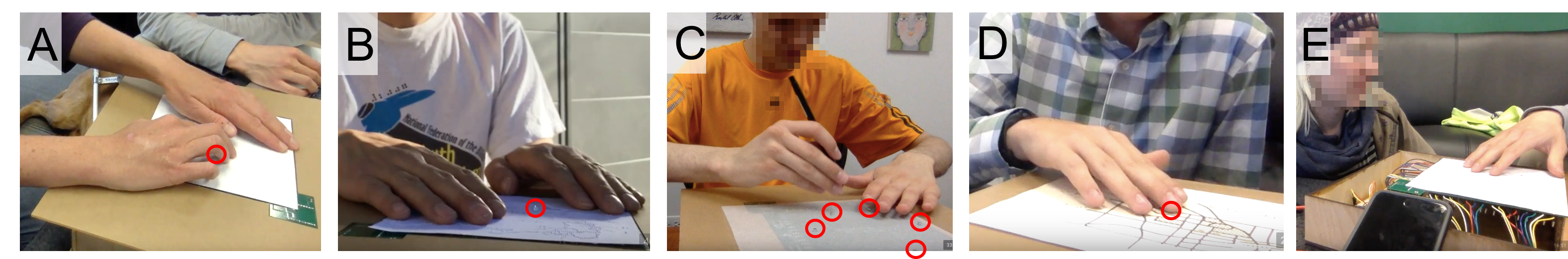}
\caption{User study. Participants used our system to (A) identify a point of interest, (B) explore a sectional view of a human brain, (C) draw a shape, and (D) navigate a street map, where red circles indicate the positions of one or more FluxMarkers. (E) A participant was shown the inner working of our system.}
~\label{fig:userstudy}
\vspace{-0.2in}
\end{figure*}

\section{User Study}
The goal of the user study was to use the prototype to assess the use case scenarios we identified during our formative research and background research, and to probe users about other possible applications of the \tool{}. In particular we observed how the tool supported participants' ability to find specific locations within a tactile graphic, supported participants ability to relate content knowledge to elements on a tactile graphic; engaged participants in drawing tasks; and affected participants perceptions of how information can be communicated with tactile graphics and their interactions with assistive technologies. 

\subsection{Participants}
Six people with visual impairments participated in the user study (3 male, 3 female); three participants were also part of the earlier formative study. 
P1 (male) P2 (male) P3 (female) identified as being totally blind. One female participant identified as being legally blind with a little bit of light perception (P4). One male (P5) and one female (P6) participant identified as having a visual impairment, but had functional vision through the use of assistive technologies. Figure~\ref{fig:participant-chart} summarizes the characteristics of participants in terms of frequency of use of tactile graphics, familiarity with science graphics, familiarity with tactile maps, and braille fluency. 
 
\subsection{Method}
We conducted a 45-minute session with each participant. During each session we presented an overview of the research, introduced the prototype and described how it worked in conjunction with the graphics, and then showed the participant two embossed tactile graphics from the local university's accessible media lab so that they would have basic familiarity with the graphics. The graphics included (A) an embossed tactile map of Eastern Europe and Russia and (B) an embossed tactile graphic representing a sectional view of a human brain. At the beginning of the session we provided the participants with the context in which these graphics might be used and provided time for them to explore the graphics. We then asked the participants to (C) draw a hexagon on a piece of trace paper, in order to observe their familiarity with drawing without any aids. Figure~\ref{fig:userstudy} shows examples of user study sessions.

To observe how the participants used the \tool{} we laid the embossed tactile graphics on top of the display and asked the participants to read graphics A and B, and perform a series of tasks with the aid of the dynamic markers. \changes{When evaluating FluxMarker with Graphic A, we asked participants to first find a region on the map without the aid of the marker, and then find a specific point on the graphic as marked with the FluxMarker. Subsequent to finding the marker, we asked participants to identify other geographic features. This allowed us to observe how each participant used the marker as a reference point throughout their search. We performed the same order of operations with Graphic B, albeit the graphic was more detailed and the representation of the brain had less "regions" and more features represented. We asked participants to identify these features in relation to each other.}We also asked participants to follow a moving marker to draw a hexagon on a piece of trace paper. While the participants were performing the task, we answered any questions that arose. We observed their actions, recorded their commentary. After these activities we conducted 10-minute semi-structured interview where we asked about their experience with the \tool{} in relation to their first and third experience with the graphics. We also asked for feedback about the prototype and their view of its current and future application. To analyze the data we reviewed the video of the sessions and captured questions and comments that arose during the testing, and identified themes that arose from the interview questions.

\subsection{Findings \& Discussion}



\subsubsection{Applications}
In order to assess the application of \tool{} we asked participants to use the tactile map, tactile graphic, and drawing paper to preform a task with the tool. Each participant performed those tasks in slightly different ways, and provided unique feedback and new ideas about the effectiveness of the tool. 

\textbf {Spatial Navigation:}
When viewing the tactile map, P1 rapidly scanned the display area with two hands and found the boundaries of the countries represented on the map without guidance; he said that he loves geography and is good at geometry. He immediately started looking for the Black Sea, at which point we used the \tool{} to help him locate the sea. He found the marker within seconds and noticed that it was positioned in the middle of the sea. P1 compared his experience with this tool as being similar to working with a teacher of the visually impaired (TVI), who might manually place a simple magnet or sticker on the map to mark a location. He suggests that we use the \tool{} to guide someone to follow a path in order to discover a landmark, in this case {\it ``the Yangtze River''} if this map also included China. At the end of the user study he said {\it ``The best application I could see this being used for is to have the marker move with the user following along, so that the teacher could trace a path out for me in real-time.''}

P3 also identified real-time mapping as an important application of the tool. She suggested {\it ``If you could have a tactile map, where then you could locate two buildings [using the markers], and then figure out pathways between those markers [which represent the buildings], you could then start populating the map with landmarks using these markers.''} P5 and P6, both low vision, explored the tactile graphic visually and did not have any ideas for how this tool would support them with navigation. 

\textbf{Feature Identification and Locating:}
In addition to using the markers to identify specific landmarks or geolocations, P3 wanted the markers to form into a raised line around specific regions of the tactile graphic to make the boundaries more amplified. When viewing the tactile map with the marker, it was located in the middle of an empty space. She asked, {\it ``I was wondering, is the marker in the middle [of the country]?''} She then suggested that the dynamic markers would be more beneficial if they could outline the boundary of the country or entity one was trying to find.

When using the dynamic markers to explore the map, P2 started brainstorming other possible applications. He provided the scenario of using Google to find restaurants with four stars, and the then using the \tool{} to automatically populating the location of the restaurants to narrow his decisions about where to go.

When using the markers to find Kazakhstan on the tactile map, P4 indicated that the markers provide a sense of independence. {\it ``It works better than having another person poking at the spot. Even if you know where their finger is, and you start taking time to explore around, they might think you are lost--which you are not--and try to show you around.``} She also mentioned that if an instructor was talking about a specific location on a graphic or map, it would be easier to keep up if the marker was in the corresponding position on the display. {\it ``This would be useful if it was synced up with a lecture and graphics, or even if it was synced with an instructors laser pointer; if it was tracking what was up on the board, and I could follow along, that would be amazing.''} 

P6 elaborated on this concept, indicating that as somebody with low vision, she has a hard time following lectures that have slides. {\it ``I usually ask the presenter to give audio cues when they are changing slides so I can follow along with the slides in front of me, but they usually forget to do that. Or if they have annotated graphics and they forget to describe that...then this tool would be very helpful. If this could be used to help track those animations on a print out of their slides. I think this would be great for low vision.''} 

\textbf{Data Analysis and Visualization:}
P3 mentioned wanting to be able to rapidly zoom into a specific area of a graphic, and see the details in higher resolution. She envisioned that the user could specify a region of a graphic, and \tool{} could dynamically represent the zoomed in region in higher resolution next to the original view. P2 suggested that \tool{} could be used to display incoming financial data from wall street to show how the market fluctuates. 

P4, an astrophysicist, remarked that while \tool{} is not yet useful for her work, she enthusiastically recommended connecting \tool{} to Excel so that she could create diagrams of her data in real-time using multiple markers. {\it ``If they [the markers] were more stable and you had an ordinary piece of tactile graphic paper, you could add the markers on top of that and make a line graph. That would be very useful. That would absolutely be helpful to me in my professional career. On of the things that happens when I am writing a paper is preparing my data in Excel, I can't get any feedback from that graph. If I had something like this plugged into my computer, I could see if it graph matches my numbers. I could actually check my own work before publishing. That is a big deal!''}

\textbf{Drawing or Tracing Guide:}
All participants found the task of following the dynamic marker to draw a hexagon to be slow and troublesome as they already had strategies for how to draw the shape. For example, when drawing a hexagon freehand, P4 first drew a square and then used her spatial understanding to add additional sides. She remarked that it was not different from following a real person and it did not provide the tactile feedback that a drawing board would. P2 remarked that the \tool{} would be difficult to use to draw organic shapes due to the orthogonal layout of the coil grid. He also remarked that it would be more useful if the markers could be closer together. However, P1 pointed out that it could provide a sense of independence for people who want to practice drawing on their own; P3 thought that the \tool{} would be a good resource for young children learning to draw. 

\subsubsection{Technical Evaluation}

The user study provided us an opportunity to evaluate the technical characteristics of the \tool{}. Here we discuss users' comments with respect to support, perceptibility, and scalability: 

\textbf{Support:} 
The design of the physical prototype enabled the user to directly place embossed tactile graphics on the PCB display. During the user study, we found that the markers moved well along embossed paper graphics, but Swell paper was too thick for the current level of magnetization to maintain contact with the display. Throughout the user study the participants remarked that the markers felt loose and that they moved too easily when touched; they wanted them to have a stronger magnetic connection to the PCB display. P4 remarked {\it `` If I was taking a test and was in a hurry looking around, and moved the marker, it would slow me down.`` We noted that in some cases this made participants hesitant to freely explore the graphics.''}



\textbf{Perceptibility:} 
P4 and P5 explicitly said something about the markers. P4 noted that the marker stands out in comparison to the rest of the page. P5 wanted to make sure that they would not cover any important graphic content. P1, P3, and P4 each remarked on the heat of the magnets and underlying PCB display. When looking for the marker P4 said, {\it ``In some ways I think the heat is a good indicator of where the marker is going to be...because the marker is a very specific spot, but the heat is a region.''} The heat she was referring to is the result of current flowing through the coil, which turns out to be a useful side-effect for P4. In contrast, P2 found the heat less favorable and noted that if we increase resolution of the markers, the heating problem would need to be addressed.

\textbf {Scalability:}
All participants were impressed by the low cost of the prototype and the prospect of a larger display size. P1 mentioned that if the resolution were lower (the markers more spread out), a mechanism would be needed to help users find their starting reference point. Without this it would be laborious to find the marker. 
P3 was satisfied with the current size of the display since she could explore the whole display in a short amount of time.

\section{Limitations \& Future Work}

Our system, while promising, has several limitations that require future research.
\changes{
First, our user study is largely based on qualitative findings, rather than quantitative measures from controlled experiments.
Although the focus of our study was to explore the application space with a proof-of-concept prototype, we acknowledge the importance of more in-depth and rigorous user study about the various aspects of the system.
For example, our system has a GUI component to enable designers to specify dynamic marker configurations, but we have not tested this formally.
For the future work, we will recruit other stakeholders including family members, teachers of visual impairments, and professional tactile transcribers to gain insight into content creation involving dynamic markers.

Second, our system currently does not detect the positions of the markers.
Instead, it assumes the initial positions of the markers are known and tracks their movements based on users' commands. 
We are interested in developing a closed-loop system which can detect markers' locations by using camera-based tracking or reed switch arrays, and then feed location information back to the system in real-time.
Such system can improve the accuracy and robustness of marker movement.

Finally, the current system uses dynamic markers as {\it output} of information, but the markers can be also used by users to provide {\it input}.
For example, one participant was excited about the potential use scenarios where he can use tactile markers to find nearby restaurants, and then he can ask context-aware questions such as open hours, ratings, or a menu of the restaurant by touching the marker, just like Google Maps.
In these scenarios, tactile markers should be enhanced to respond to real-time information requests.
For the future work, we are interested in integrating touch sensing or gesture tracking mechanisms into the system to open up a new interaction model for visually impaired users and explore the possible design space of dynamic tactile markers.
}



\section{Acknowledgments}
We would like to thank the participants for their time and helpful feedback.
We also would like to thank Shohei Aoki and Ron Pelrine for their technical advice and feedback.
This research was supported by the NSF CAREER award 1453771 and the Nakajima Foundation.

\balance

\bibliographystyle{abbrv}
\bibliography{references}

\end{document}